\documentclass[prl,twocolumn,superscriptaddress]{revtex4}

\usepackage{amsmath,amsfonts,amssymb,amstext,amsthm}
\usepackage{graphicx}
\usepackage[colorlinks,citecolor=blue]{hyperref}

\def\>{\rangle}
\def\<{\langle}
\def\({\left(}
\def\){\right)}

\newcommand{\ket}[1]{|#1\>}
\newcommand{\bra}[1]{\<#1|}
\newcommand{\braket}[2]{\<#1|#2\>}
\newcommand{\ketbra}[2]{|#1\>\!\<#2|}

\newcommand{\be}{\begin{equation}}
\newcommand{\ee}{\end{equation}}
\newcommand{\eref}[1]{Eq.~(\ref{#1})}

\def\bbC{\mathbb{C}}

\def\bbZ{\mathbb{Z}}

\DeclareMathOperator{\Tr}{Tr}

\def\r{R\'{e}nyi }

\def\bbbone{{\mathchoice {\rm 1\mskip-4mu l} {\rm 1\mskip-4mu l}
{\rm 1\mskip-4.5mu l} {\rm 1\mskip-5mu l}}}

\begin{document}

\title{Topological Entanglement \r Entropy and Reduced Density Matrix Structure}

\author{Steven T. Flammia}
\author{Alioscia Hamma}
\affiliation{Perimeter Institute for Theoretical Physics, Waterloo, Ontario, N2L 2Y5 Canada}

\author{Taylor L. Hughes}
\affiliation{Department of Physics, Stanford University, Stanford,
CA, 94305}
\affiliation{Department of Physics, University of Illinois, Urbana-Champaign, Urbana,IL,61802}

\author{Xiao-Gang Wen}
\affiliation{Department of Physics, Massachusetts Institute of Technology, Cambridge, Massachusetts 02139}
\affiliation{Perimeter Institute for Theoretical Physics, Waterloo, Ontario, N2L 2Y5 Canada}

\date{December 1, 2009}

\begin{abstract}
We generalize the topological entanglement entropy to a family of
topological \r entropies parametrized by a parameter $\alpha$, in an
attempt to find new invariants for distinguishing topologically
ordered phases.  We show that, surprisingly, all topological \r
entropies are the same, independent of $\alpha$ for all non-chiral topological phases.  
This independence shows that topologically ordered ground-state wavefunctions have reduced density matrices with a certain simple structure, and no additional universal
information can be extracted from the entanglement spectrum.
\end{abstract}

\maketitle

\paragraph{Introduction.}
Topological order (TO) \cite{wenbook} is a new kind of order that
corresponds to patterns of long range quantum entanglement which
cannot be described by symmetry breaking.  However, the long range
quantum entanglement in TO can leave its mark on the reduced
density matrix, so one may be able to study long range
entanglement and TO through the structure of these reduced
correlations. The reduced density matrix contains a lot of local
non-universal information. The key is to filter out all the
non-universal information to capture the universal topological
information, which is not affected by perturbations of the
Hamiltonian, or small deformations of the entanglement partition
geometry.  One way is to calculate \emph{topological}
entanglement entropy (EE) from reduced density matrices
\cite{Hamma2005, Kitaev2006a, Levin2006,Haque2007,Furukawa2007}.  Such a universal
quantity
provides a way to determine whether or not a ground state
possesses TO. If we only consider systems with a finite excitation
gap, the low-energy physics can be described in terms of an
underlying topological quantum field theory (TQFT). Then the
topological EE is proportional to the logarithm of the total
quantum dimension $S_{\rm top}\propto \log D$. Unfortunately the
quantum dimension does not provide a complete classification of
TO. For example, two topologically ordered states, the $\bbZ_2$
gauge theory and Ising anyons \cite{Read2000,Kitaev2006,Yao2007a},
are different phases of matter --- with 
Abelian and non-Abelian anyonic excitations, respectively.  However, they have the same $S_{\rm top}
= \log 2$.  To obtain a finer classification of TO, Ref.
\cite{Li2008a} proposes using the entire entanglement spectrum (possibly with additional conserved quantum numbers.)

These developments motivated us to consider an approach which might
glean more universal information from the entanglement spectrum.
We introduce a generalization of the topological EE by deforming
it into a \r entropy parameterized by a real number $\alpha$ which
can characterize different aspects of the entanglement spectrum
akin to moments of a probability distribution. We calculate this
entropic quantity for the exactly solvable string-net
\cite{Levin2005a} and quantum double \cite{Bais1992,Kitaev2003a}
models, which describe all the non-chiral topological phases. Recent works have mapped the quantum double models onto a
subset of string-net models \cite{Buerschaper2009, Kadar2009}, so
we can compare entropies calculated for two different
wavefunctions with the same TO.  Our central result is that the
only universal information captured by the \r entropy is the
quantum dimension $D$, {\it i.e.} the topological \r entropy does
not depend on the extra parameter $\alpha$.  As a consequence, no
more universal information about the TO phases can be extracted
from the entanglement spectrum without additional conserved quantum numbers. 
Such a result suggests that the
reduced density matrix $\rho_A$ for a subregion $A$ formally has
the following structure $\otimes \rho_i=\rho_A \otimes
\rho_\text{top}$, where $\otimes \rho_i$ is the tensor product of
the local density matrices of the degrees of freedom living on the
boundary of $A$.  The ``topological'' desity matrix
$\rho_\text{top}$ has a simple form where all its non-zero
eigenvalues are equal, which leads to the $\alpha$ independence of the 
topological \r entropy, which we demonstrate explicitly for the quantum double models.


\paragraph{\r entropy.} The quantum \r entropy is defined with respect
to a parameter $\alpha > 0$ as
\begin{align}\label{E:renyi}
    S_\alpha(\rho) = \frac{1}{1-\alpha} \log\Big[\Tr\!\big(\rho^\alpha\big)\Big] \, ,
\end{align}
where the base of the logarithm is chosen to fix the units with which
one measures the entropy.  Taking the limit as $\alpha \to 1$, one
recovers the definition of the von Neumann entropy $    \lim_{\alpha
\to 1} S_\alpha(\rho) = S_1(\rho) = -\Tr\!\big(\rho \log \rho\big) $.
The \r entropy is additive on independent states in the sense that the
entropy of a product state is the sum of the individual entropies, $
S_\alpha(\rho \otimes \sigma) = S_\alpha(\rho) + S_\alpha(\sigma)$.
The \r entropy is essentially unique if we look for a function that is
symmetric, continuous, has the additive property, depends only on the
spectrum of $\rho$, and obeys a generalized mean value property
\cite{Renyi1961}.  This essential uniqueness given certain natural
assumptions and desired properties, together with the fact that the \r
entropies cover a very broad class of functions motivates their
consideration as a classification tool for TO.


\paragraph{String-net states.} We can study all parity-invariant
topological phases in $(2+1)$-d using string-net models
\cite{Levin2005a}. These models exhibit TO and represent an
exactly solvable fixed point in a topological phase. The degrees
of freedom are a set of strings living on the links of a honeycomb
lattice. To specify a string net model requires several
ingredients: a set of $N$ string types $i=1,\ldots, N$, a
branching rule tensor $N^i_{jk},$ and two real tensors $d_{i}$ and
$F^{k\ell m}_{hij}$  which satisfy certain algebraic relations
\cite{Levin2005a} to ensure consistency. Every string type $i$ has
an oppositely oriented partner $\bar i$. The ground state
wavefunctions of the string-net models obey a concise set of
diagrammatic rules which are characterized by the string-net data
listed above. In Ref.~\cite{Levin2006} the (von Neumann, $\alpha
\to1$) topological EE for such string-net models was defined and
calculated to be $S_{\rm top} = \log D^2$ where the quantum
dimension $D=\sum^{N}_{i=1}d^2_{i}.$ Thus, from a knowledge of the
ground state one can extract universal information about the
low-energy TQFT and underlying TO in the form of the total quantum
dimension.


\paragraph{\r entropy for string-nets.}  To define an EE we begin by partitioning our system into two pieces. In this
letter we will focus on a simply connected region A and trace out its
exterior. The region A is topologically a disk and the reduced density
operator of the string-net model on the disk can be deformed into a
sum over string configurations on a tree-like diagram at the boundary
of the disk \cite{Levin2006}.  We assume that our boundary string-net
tree diagram has $n$ boundary nodes with  $n$ links of the boundary
tree labelled $q_i$ connected by $n-3$ internal links.  To begin the
\r entropy calculation we start from Eq. 9 in Ref.~\cite{Levin2006},
which gives the reduced density operator in region $A$, which we label
by $\rho_A$. We first raise $\rho_A$ to the power $\alpha$ and trace,
summing over the states by using the branching rules $N^i_{jk}$ to get
\begin{align}\label{E:trrho}
    \Tr(\rho_A^\alpha) = \frac{D^\alpha}{D^{\alpha n}} \sum_{\{q\}} N_{\{q\}} \prod_m d_{q_m}^\alpha \, ,
\end{align}
where the expression for $N_{\{q\}}$ is given succinctly in terms of
the matrices $\hat N_q = \sum_{a,b} N_{a q}^b \ketbra{a}{b}$, whose
basis states form an orthonormal basis labelled by the string types: $
N_{\{q\}} = \bra{q_1} \hat N_{q_2} \hat N_{q_3} \cdots \hat
N_{q_{n-1}} \ket{q_n}$. By relabelling the boundary strings in terms
of the real-valued vector $\ket{d^\alpha} = \sum_q d_q^\alpha \ket q$,
we can return to \eref{E:trrho} and write
\begin{align}
    \Tr(\rho_A^\alpha) = \frac{D^\alpha}{D^{\alpha n}} \sum_{\{q\}} \bra{d^\alpha} \hat N_{q_1} d_{q_1}^\alpha \cdots \hat N_{q_{n-2}} d_{q_{n-2}}^\alpha \ket{d^\alpha} \, ,
\end{align}
where the sum on $\{q\}$ runs only over $n-2$ different $q_i$.  Since
we are summing over all possible combinations, we can collect terms to
get the even simpler form
\begin{align}
    \Tr(\rho_A^\alpha) = \frac{D^\alpha}{D^{\alpha n}} \bra{d^\alpha} \(\sum_q \hat N_q d_q^\alpha\)^{n-2} \ket{d^\alpha} \, .
\end{align}
We can make use of some properties of the $\hat N_q$ matrices to
simplify this expression. The $\hat N_q$ satisfy $\hat N_q^\dag = \hat
N_{\bar q}$ (where $\bar q$ annihilates $q$) and if braiding is
defined, we have $N_{ab}^c = N_{ba}^c$, which implies that all the
$\hat N_q$ commute with each other.  This means that the $\hat N_q$
are normal and can be unitarily diagonalized simultaneously.  Let $S$
be the matrix such that  $S^\dag \hat N_q S = \Lambda_q$ is diagonal.
Then we also have $\sum_q \hat N_q d_q^\alpha = S (\sum_q \Lambda_q
d_q^\alpha) S^\dag$.

Under the additional assumption that the braiding is sufficiently
nontrivial (as discussed in the Appendix of Ref.  \cite{Kitaev2006}),
we have so-called modularity, and the $S$ described above is indeed
the unitary modular $S$-matrix of the theory. We choose the S-matrix
to be in the canonical form where we can read off the quantum
dimensions from the first row or column. As we will see, this puts the
largest eigenvalue of $\sum_q \Lambda_q$ in the first matrix element.

Since the $\hat N_q$ are normal and mutually commuting, they share in
common a complete set of orthogonal eigenvectors.  Each $\hat N_q$ has
an eigenvalue $d_q$ with the eigenvector $\ket d$.  Moreover, due to
the Perron-Frobenius theorem, every other eigenvalue $\lambda$ for
each $\hat N_q$ satisfies $|\lambda| \le d_q$.  Thus we know exactly
what the largest eigenvalue of $\sum_q \hat N_q$ is, namely $\sum_q
d_q$.  For symmetric matrices (and $\sum_q \hat N_q$ is symmetric),
the Perron-Frobenius theorem gives us additional guarantees.  In
particular, the largest eigenvalue $\lambda_{\rm max}$ is
non-degenerate.  Furthermore, the least eigenvalue satisfies
$\lambda_{\rm min} = - \lambda_{\rm max}$ if and only if the symmetric
matrix is the adjacency matrix of a bipartite graph.  But this can't
be the case, since the vacuum always fuses with itself to form the
vacuum, giving at least one nonzero element on the main diagonal, and
bipartite graphs have no self-loops.  Therefore all other eigenvalues
$\lambda$ of $\sum_q \hat N_q$ satisfy $|\lambda| < \sum_q d_q$, and
these $\lambda$ contribute exponentially less once we raise to the
power $n-2$.  Then, ignoring a multiplicative factor of $(1+
O(\exp(-n))),$ we have
\begin{align}\label{E:lastrenyi}
    \Tr(\rho_A^\alpha) = \frac{D^\alpha |\bra{d^\alpha} S \ket{1}|^2}{D^{\alpha n}} \braket{d^\alpha}{d}^{n-2} \, .
\end{align}

To get a more explicit expression, we need to calculate
$|\bra{d^\alpha} S \ket{1}|^2$. Let's consider how $S$ acts on
$\ket 1$.  $S$ is a unitary matrix, and the first row is
proportional to $\bra d$.  So $S \ket 1 = \frac{1}{\sqrt{D}}\ket
d$.  Hence $|\bra{d^\alpha} S \ket{1}|^2 = \braket{d^\alpha}{d}^2/D$, and
substituting this in to Eq.~(\ref{E:lastrenyi}) and using the expression for the \r entropy in Eq.~(\ref{E:renyi}), we obtain
\begin{eqnarray}
S_{\alpha}(\rho_A) = \frac{n}{1-\alpha}\log
\(\frac{\braket{d^\alpha}{d}}{D^\alpha}\)-\log D ,
\end{eqnarray}
which is correct up to a term of order $O(\exp (-n))$.  The first
term represents the area law. It is not universal and cannot be
used to describe the phases.  The second term represents the
universal part: the topological entanglement \r entropy.  We see
that it does not contain any $\alpha$ dependence, just the total
quantum dimension $D$.  Therefore the it does not provide any
additional universal information. The \r entropies completely determine the
spectrum, hence no additional information (beyond $D$) can be
gathered from the entanglement spectrum. This is true when the
partition geometry is simply connected; Ref. \cite{Dong2008a} has
shown that more can be extracted in more complicated partitions.

We wish to find deeper insight into why there is nothing else in
the eigenvalues of the reduced density matrix that can say more
about topological order. To this end, we will study an important
class of TO states, those emerging from discrete gauge theories.
In the following, we prove that the reduced density matrix of such
states is proportional to a projector, and thus
all \r entropies contain no $\alpha$ dependence and that the whole
entanglement spectrum is trivial and flat.


\paragraph{Quantum Double Models.} 
The quantum double models are
exactly solvable lattice models with discrete gauge
symmetries~\cite{Bais1992,Kitaev2003a}. These models exhibit
phases with TO and anyonic excitations, and are in the same
universality class as a subset of the string-net
models~\cite{Buerschaper2009}. To define them, begin with a
directed graph with orientations $\pm$ and with qudits on the
edges.  Consider a finite group $G$ of dimension $|G|= d$, with
identity $e$. The local Hilbert space on the edge $i$ is therefore
$\mathcal H_i  \simeq \bbC[G]$ and an orthonormal basis for the
qudits is given by $\{\ket{g}:g\in G\}$.  The total Hilbert space
for a system with $n$ qudits is given by $\mathcal H =
\otimes_{i=1}^n \mathcal H_i$.  We focus on the model on a square
lattice, with $n/2$ vertices and plaquettes.

Following the construction of~\cite{Kitaev2003a}, the relevant
operators are $L^g_{\pm}, T^h_ \pm$ defined by $L^g_+\ket{z}=
\ket{gz}, T^h_+\ket{z} = \delta_{h,z}\ket{z}, L^g_-\ket{z}=
\ket{zg^{-1}}, T^h_-\ket{z} = \delta_{h^{-1},z}\ket{z}$. The gauge
transformations are defined as follows:
\begin{align}
    A_g(s) = \prod_{j\in s} L^g(j,s) \ ,\ B_e (p)= \hspace{-6pt} \sum_{h_1h_2h_3h_4 = e}\prod_{m=1}^4 T^{h_m}(j_m,p). \nonumber
\end{align}
The star and plaquette operators are defined as the projector
operators $A(s) =|G|^{-1}\sum_{g \in G} A_g(s), B(p) = B_e(p)$.
The Hamiltonian of the quantum double model is
\begin{equation}
H_{QD}  =  \sum_s \left( 1- A(s)\right) +\sum_p\left(1-B(p)\right)
\end{equation}
Since $[A(s),B(p)]=[A(s),A(s')]=[B(p),B(p')]$ for all $s,s',p,p'$, the
ground state manifold is given by the set
$\mathcal L
=\{ \ket \xi \in \mathcal H \ | \ A(s)\ket \xi =B(p)\ket \xi = \ket
\xi \ \forall s,p \}
$
with ground state energy $E_0=0$.

Consider the vacuum state $\ket{e} = \ket{e}^{\otimes n}$.  For
each plaquette $p$, it easily follows that $B(p)\ket{e} =
\ket{e}$.  We can build a (un-normalized) ground state
$\ket{\xi_0} \in \mathcal L$ by projecting as follows,
$\ket{\xi_0} = \prod_s A(s)\ket{e}$.

Now, consider the set $\mathfrak G$ of all the possible $A_g(s)$. What
this operator does is to make a small loop around $s$ with string of
type $g$. We have $\mathfrak G = \{  A_g(s), g\in G, s=1,...,n/2
\}$. Now consider the set $\mathcal G = \langle \mathfrak G\rangle$,
that is the set of all the possible products of elements in $
\mathfrak G$. The set $\mathcal G$ is a group. With this definition,
we have
\begin{equation}\label{gs} \ket{\xi_0} = |G|^{-1}\prod_{s}
\sum_{g \in G} A_g(s)\ket{e}=|G|^{-\frac{n}{2}}\sum_{h\in \mathcal
G}h\ket{e}
\end{equation}
It is important to see that the set of $\{\ket{h}\}$ is
orthonormal. Moreover, given a bipartition of the Hilbert space
$\mathcal H = \mathcal H_A \otimes \mathcal H_B$, the set
$\{\ket{h_A}\otimes\ket{h_B}\}$ is bi-orthonormal.  Let us compute
the density matrix $\rho_0 = \ket{\xi_0}\bra{\xi_0}$. Since each
vector $\ket h$ factorizes as $\ket{h_A} \otimes \ket{h_B}$, we
have
\begin{equation} \rho_0 = |G|^{-n}  \sum_{h,h' \in \mathcal G}
\ket{h_A}\bra{h_A'} \otimes  \ket{h_B}\bra{h_B'} ,
\end{equation}
Consider now the subgroup of $\mathcal G$ acting exclusively on
subsystem $A$, $\mathcal G_A := \{g\in \mathcal G\ |\ \  g= g_A\otimes
\bbbone_B\}$, and analogously consider $\mathcal G_B$.  It is easy to
show that  $\mathcal G_A$, $\mathcal G_B$, and $\mathcal
G_A\times\mathcal G_B$ are {\em normal} in $\mathcal G$.  Therefore we
can define the quotient groups $\mathcal G_{AB}:= \mathcal G/\mathcal
G_A\times \mathcal  G_B,\mathcal G/\mathcal G_B,\mathcal G/\mathcal
G_A$. We see that the only elements of $\mathcal G$ such that
$\bra{e}\tilde{h}_B\ket{e}\ne 0$ are those in $\mathcal G_A$, and
therefore we find
$ \rho_A = |G|^{-n}  \sum_{h\in
\mathcal G,\tilde{h}\in \mathcal G_A} \ket{h_A}\bra{h^{-1}_A
\tilde{h}_A} $, where we have relabelled the group elements as $h' =
h^{-1}\tilde{h}$. Notice that $ \ket{h_A}\bra{\tilde{h}_A} = g
\ket{h_A}\bra{\tilde{h}_A}$ for every $g\in \mathcal G_B$, and
$|\mathcal G| =|G|^n$. Therefore, reordering gives
\begin{eqnarray}
\rho_A &=&  |\mathcal G|^{-1}|\mathcal G_B| \sum_{h\in \mathcal
G/\mathcal G_B,\tilde{h}            \in \mathcal G_A}
\ket{h^{-1}_A}\bra{h_A \tilde{h}_A}
\end{eqnarray}
Squaring this expression for $\rho_A$ and using the group
properties shows that $\rho_A$ is proportional to a projector,
$\rho_A^2 = \frac{|\mathcal G_A| | \mathcal G_B|}{|\mathcal G|}
\rho_A$, and therefore the \r entropies contain no additional
information beyond the quantum dimension, $D = |G|$. The
entanglement spectrum is flat which is connected \cite{Li2008a}
with the trivial nature of the edge states for the QD models.


\paragraph{The origin of the topological term.} 
At this point, we would
like to understand {\em why} $\rho_A$ is just a projector?  And
why, in the more general string-net setting where the reduced
density matrix is not just a projector, is there still no
topological information other than $D$?
Here we prove that the reduced density matrix $\rho_A$ is
unitarily equivalent to a matrix that only addresses the degrees
of freedom on the boundary of the partition. Moreover, we show
that the area law has a correction because there is a global
constraint on the boundary. We can enlarge the system by removing
this constraint and express the reduced density matrix as the
tensor product of the local density matrix of each of the degrees
of freedom on the boundary. We focus on the $\bbZ_2$ case for
simplicity, but the argument can be generalized to all the quantum
double models. In this case, the ground state is given by
Eq.~(\ref{gs}), where $\mathcal G$ is the group generated by the
plaquette operators $A_p = \prod_{j\in
\partial p} \sigma^x_i$ and $\ket{0}$ is the state with all spins up
in the $z$-basis.  By choosing a simply connected region of
plaquettes, we partition the spins into $(A,B)$, where $A$
includes the spins in the interior and on the boundary. The
quotient group $\mathcal G_{AB}$ consists of the closed strings
that act on both $A$ and $B$, that are equivalent under 
deformations acting entirely within $A$ or
$B$. Therefore, the equivalence classes in $\mathcal G_{AB}$ can
be represented by those closed strings that live near the boundary
between $A$ and $B$, namely those closed strings that are
generated by the plaquettes that are external to $A$ and share one
edge with the boundary. So every element $h\in \mathcal G_{AB}$
can be decomposed as $h = h_A\otimes h_B$ where $h_A$ only acts on
spins that live on the boundary (``$a$'' spins). The $h_B$ part
only acts on those spins which are external to $A$ (``$b$''
spins). The rest of the lattice consists of the spins in the bulk
of $A$ and $B$, namely all those spins that belong solely to
either $A$ or $B$: $\ket{0} = \ket{0}_{a}\otimes \ket{0}_b
\otimes\ket{0}_{\rm bulk}$ so that $h\ket{0} = h_A\ket{0}_a\otimes
h_B\ket{0}_b\otimes I\ket{0}_{\rm bulk}$. Therefore the ground state
can be written as \be\nonumber \ket{\psi} = |\mathcal G|^{-1/2}
\hspace{-12pt} \sum_{\substack{g_A\otimes g_B \in \mathcal
G_A\times \mathcal G_B
\\ h\in \mathcal G_{AB}}} \hspace{-10pt} h_A\ket{0}_a\otimes
h_B\ket{0}_b\otimes (g_A\otimes g_B) \ket{0}_{\rm bulk} \ee Define
$Q_X = |\mathcal G_X|^{-1/2}\sum_{g_X\in \mathcal G_X} g_X$, with
$X=A,B$. We obtain $\ket{\psi} =|\mathcal
G_{AB}|^{-1/2} Q_AQ_B\sum_{h\in \mathcal G_{AB}}h_A\ket{0}_a\otimes
h_B\ket{0}_b\otimes \ket{0}_{\rm bulk}$. The density matrix
can be therefore be factored as $\rho \equiv
Q_AQ_B \tilde{\rho}\otimes \rho^{\rm (bulk)} Q_AQ_B$, with
\begin{eqnarray}
\nonumber \tilde{\rho}\otimes \rho^{\rm (bulk)} = |\mathcal G_{AB}|^{-1}\sum_{h,h'\in \mathcal
G_{AB}}h_A\ket{0} \bra{0}_a h'_A\otimes h_B\ket{0}\bra{0}_b  h_B' \\ \nonumber
 \otimes \ket{0}\bra{0}_{\rm bulk} \, .
\end{eqnarray}
Notice that the bulk part is separable in the bipartition $(A,B)$
so that the reduced density matrix can be written as $\rho_A =
Q_A \tilde{\rho}_A  \otimes \rho^{\rm (bulk)}_A Q_A$, with $\rho^{\rm (bulk)}_A $
being a pure state, so the EE of $\rho$ is just the
entropy of $\tilde{\rho}_A$. Then
\be\label{red}
	 \tilde{\rho}_A =|\mathcal G_{AB}|^{-1} \sum_{h\in \mathcal G_{AB}}h_A\ket{0} \bra{0}_a h_A \,,
\ee 
which gives the
expected result for the entanglement $S_1(\rho) = \log _2
|\mathcal G_{AB}|$ \cite{Hamma2005,Hamma2005a}. We are interested in
understanding what are the spin configurations in the sum
Eq.~(\ref{red}). Notice that the support of $\tilde{\rho}_A$
consists only of the spins on the boundary.  Now, note that every
spin configuration is not allowed. In fact, we have the following
global constraint: $\prod_{h\in \mathcal G_{AB}} h = g_A\otimes
g_B \in \mathcal G_A\times \mathcal G_B$. So the product of all
the $h_A$ is also in $\mathcal G_A$ and is $+1$ on the ground
state. The global constraint is thus $\prod_{j \in
\partial A} \sigma^z_j = +1$ so that the reduced density matrix
Eq.~(\ref{red}) consists of the sum of all spin configurations with
parity $+1$, namely all the spin configurations with an even pair
of spins flipped. If the boundary has length $n$, then there are
$2^{n-1}$ such configurations and the entanglement is then $S=
n-1$. Now we understand that the topological state is completely
determined by the boundary, and that we have the completely mixed
state within the sector of parity $+1$. We can consider the
enlarged system by considering the perfect mixture with the sector
of parity $-1$. In this case, we have that
\begin{align} \rho^{\rm (area)}_A
= \tilde{\rho}_A \otimes \left( \begin{array}{cc} 1 &0 \\ 0
&1\end{array}\right)\equiv  \tilde{\rho}_A \otimes \rho_A ^{\rm (top)} .
\end{align}
So $\rho^{\rm (area)}_A $ is just the completely mixed state of all
the possible spin configurations on the boundary and thus
\begin{align}
\label{rrr}
\rho^{\rm (area)}_A = \otimes_{j=1}^n \rho_j = \tilde{\rho}_A \otimes
\rho_A ^{\rm (top)} .
\end{align}
We have shown that the entanglement in the ground state of a 
topologically ordered system
is completely contained in the boundary, namely in the entropy of
the reduced density matrix $\tilde{\rho}_A$. We have also shown
that this state \emph{almost} obeys an area law, because there is
a global topological constraint, namely that only spin
configurations of parity $+1$ are allowed. Therefore, we can
complete it with a density matrix that describes a system before
we project onto the system with parity $+1$ . This term contains
the topological entropy. Once completed, the system obeys a strict
area law and decomposes into the local tensor product of the
single degrees of freedom on the boundary. Such structure of the 
reduced density matrix, as described in Eq.~(\ref{rrr}), explains
why the topological \r entropies do not depend on $\alpha$.

Finally, we remark that our proof applies to non-chiral topological phases. Therefore it is still an open problem to what extent the entanglement spectrum can classify chiral topological phases \cite{Li2008a, haas}. 

\paragraph{Acknowledgments.} We acknowledge useful conversations with 
A. Kitaev, X.-L. Qi, and H. Yao. STF, AH and XGW are
supported by the Government of Canada through Industry Canada and by
the Province of Ontario through the Ministry of Research~\&
Innovation. XGW is also supported by NSF Grant No.  DMR-0706078.


\end{document}